\documentclass[aps,pra,floatfix,superscriptaddress,showpacs,twocolumn,nofootinbib,longbibliography]{revtex4-1}
\pdfoutput=1
\usepackage{endnotes}
\usepackage{amsfonts}
\usepackage{amsmath}
\usepackage{amssymb}
\usepackage{mathtools}
\usepackage{graphicx}
\usepackage{color}
\usepackage{changes}
\usepackage{bbm}
\usepackage{physics}
\usepackage[hidelinks]{hyperref}
\hypersetup{
     colorlinks   = true,
     citecolor    = blue,
     linkcolor    = blue
}
\usepackage{matlab-prettifier}
\usepackage{ulem}

\newtheorem{theorem}{Theorem}

\newcommand{\be}{\begin{equation}}
\newcommand{\ee}{\end{equation}}
\newcommand{\bea}{\begin{eqnarray}}
\newcommand{\eea}{\end{eqnarray}}
\newcommand{\ba}{\begin{align}}
\newcommand{\ea}{\end{align}}
\def\({\left(}
\def\){\right)}
\def\[{\left[}
\def\]{\right]}

\def\6{\partial}

\def\det{\textrm{det}}

\usepackage{calligra}
\DeclareMathAlphabet{\mathcalligra}{T1}{calligra}{m}{n}
\DeclareFontShape{T1}{calligra}{m}{n}{<->s*[2.2]callig15}{}

\begin{document}

\title{Worldsheet from worldline}

\author{Umut G\"ursoy,}
\author{Guim Planella Planas}
\affiliation{Institute for Theoretical Physics, Utrecht University, 3584 CE Utrecht, The Netherlands}
\begin{abstract}
We take a step toward a ``microscopic'' derivation of gauge-string duality. In particular, using mathematical techniques of Strebel differentials and discrete exterior calculus, we obtain bosonic string worldsheet action for string embedded in d+1 dimensional asymptotically AdS space from multi-loop Feynman graphs of quantum field theory of scalar matrices in d-dimensions in the limit of diverging loop number. Our work is building on the program started by 't Hooft in 1974 \cite{tHooft:1973alw}, this time including the holographic dimension which we show to emerge from the continuum of Schwinger parameters of Feynman diagrams.  
\end{abstract}
\maketitle
\flushbottom

\section{Introduction}

Despite being widely accepted and having passed many non-trivial tests,  gauge-gravity duality \cite{Maldacena:1997re,Gubser:1998bc,Witten:1998qj} lacks a satisfactory derivation as an equivalence between quantum gauge theory and string theory. In particular, apart from certain low dimensional examples and integrable class of quantum field theories, e.g. \cite{Kazakov:1985ea,Ginsparg:1993is,Eberhardt:2018ouy,Gaberdiel:2012uj,Gaberdiel:2010pz}, we are unaware of any ``microscopic'' derivation of the string world sheet directly from the gauge theory path integral. This derivation would entail, in a concrete sense, completing the program started by `t Hooft's seminal paper \cite{tHooft:1973alw}, that is, to derive the string path integral --- this time including the holographic dimension --- from the ribboned Feynman graphs of gauge theories.  

A promising starting point for this endeavor is Gopakumar's observation \cite{Gopakumar:2003ns,Gopakumar:2004qb,Gopakumar:2005fx}. See \cite{Aharony:2020omh,Alday:2023jdk,Douglas:2010rc,Lee:2013dln,Heemskerk:2009pn,Razamat:2009mc,Ooguri:2002gx,Eberhardt:2018ouy,Berkovits:2007rj,Bargheer:2018jvq} for an incomplete list of other examples and alternative approaches. In particular, \cite{Gopakumar:2005fx} proposed to use Strebel differentials to map the skeleton graphs of n-point functions of gauge invariant operators in {\em free} CFTs to the worldsheet of a putative string theory, see also \cite{Gopakumar:2022djw,Gaberdiel:2020ycd}. Furthermore, in \cite{DomingoGallegos:2022ttp,Gallegos:2022hbt} a curious similarity between (discretized) open and closed strings and a graphical representation of one- and two-trees that appear in Symanzik polynomial representation of high-loop Feynman graphs, see e.g. \cite{Lam:1969xk,Bogner:2010kv,Weinzierl:2022eaz,Dai:2006vj,Gallegos:2022hbt}, were observed and it was proposed that embedding of the sting in the holographic dimension arises from a continuum limit of the Schwinger parameters of graphs with large number of loops. 

We merge these observations in this work and explicitly construct the string worldsheet action that arises from the continuum limit of Feynman graphs of a quantum field theory in d-dimensions. We show that the target space of the string that correspond to the QFTs we consider in this work is asymptotically AdS$_{d+1}$. 

\section{Generating function}
Consider expansion of a QFT generating function in Feynman diagrams. We restrict ourselves to QFTs of N$\times$N matrices that are massless scalars with a generic polynomial type interaction. Contribution of a graph $\Gamma$ is given by
\begin{equation}\label{eq::genrf}
\mathcal{Z}_\Gamma \propto \int \prod_i d^d X_ i \prod_{e_{ij}\in \Gamma} G(X_i,X_j)\prod_{v_i\in \mathcal{S}_ \Gamma} J(X_i)
\end{equation}
with the proportionality constant including all factors of the couplings and symmetry factors, $e_{ij}$ and $v_i$ denoting the edges and the vertices and $J$ denoting the sources. We assume, following \cite{tHooft:1973alw}, that Feynman graphs are put in ribboned form and grouped as discretized string worldsheets with genus $g$. We first rewrite (\ref{eq::genrf}) in a form that allows us to read off the corresponding (discretized) worldsheet action. 

A key role is played by the Schwinger parametrization of the propagators $G(X_i-X_j)$ with Schwinger parameters $\sigma_{ij}$. If there are $m$ edges between a pair of vertices then we can merge them in an effective propagator \cite{Gopakumar:2004qb} written in terms of a single Schwinger parameter as 
\begin{align}
\label{eq:factorsmultiplicity}
G(X_i,X_j)^m =& \frac{i^m\Gamma \left( \frac{d}{2}-1 \right)^m }{2^{\frac{m d}{2}} \Gamma \left( m \left( \frac{d}{2}-1 \right)  \right) }\cdot\nonumber\\&\cdot\int_0^\infty d \sigma_{ij} \,\sigma_{ij}^{m \left( \frac{d }{2}-1 \right)-1} e^{- \frac{\sigma_{ij} (X_i-X_j)^2}{4} }\, .
\end{align}
We assume vanishing masses throughout this paper---see \cite{Gallegos:2022hbt} for generalization of this expression to massive scalars.  This procedure constructs the so-called ``skeleton graph'' $\bar{\Gamma}$ of $\Gamma$ \cite{Gopakumar:2004qb} with all duplicated edges fused into a single one and each skeleton edge assigned a length given by the corresponding $\sigma_{ij}$. The partition function is then expressed as
\begin{equation}
\label{eq:partitionfunctiongraphs}
	\mathcal{Z}[J]=\sum_{\bar\Gamma} \int d [X]d  [\sigma]\mathcal{P}_{\bar\Gamma} e^{- \frac{1}{4}\sum \sigma_{ij} (X_i-X_j)^2}
\end{equation}
where $[X]$ and $[\sigma]$ denote the collection of all vertex positions and graph metrics of the skeleton graph and $\mathcal{P}_{\bar\Gamma}$ is a polynomial in $\sigma_{ij}$ which depends on the graph structure,  coupling constants and sources. 

To determine it, we introduce auxiliary variables $\xi_i$ at the vertices $v_i$ and represent each edge connected to a vertex by $i \xi_i$ so that for each propagator between $v_i$ and $v_j$ we must include a factor of $-\xi_i \xi_j$. 
It is straightforward to show that
\begin{align}\label{PGamma}
&\mathcal{P}_{\bar\Gamma}= \frac{1}{n!}\left(i \prod_{i} U\left( -i \partial_{\xi_i} \right)\right)\prod_{\bar{e}_{ij}\in \bar{\Gamma}} \sum_ m \tfrac{(-1)^m\Gamma\left(\frac{d}{2}-1\right)^m}{2^{\frac{m d}{2}} \Gamma \left(  m \left( \frac{d }{2}-1 \right) \right) }\cdot\nonumber\\&\left.\cdot \sigma_{ij}^{m \left(  \frac{d }{2}-1 \right)-1} \frac{(\Tr (\xi_i^{\dagger} \xi_ j + \xi_j^{\dagger}\xi_i))^m}{2 (m!)} \right|_{\xi=0}\, 
\end{align}
gives a non-vanishing contribution only when the number of edges connected to the vertex $v_i$ is precisely equal to the number of derivatives and only when the coordination number of the vertices equal to the one on some diagram $\Gamma$ whose skeleton diagram is $\bar{\Gamma}$. The symmetry factors of the graph are correctly accounted for by the factors of $1/n!$ and $1/m!$ in this expression. The latter arises from swapping of propagators between the same pair of vertices and the former from swapping of vertices. This gives the correct factor as long as we sum over all nonequivalent permutations of vertices. This sum will be implemented below as part of the integral over the moduli space, which should now be the moduli space of Riemann surfaces with $n$ marked points counting separately all nonequivalent permutations of these points. 

We recognize the exponential in (\ref{eq:partitionfunctiongraphs}) as a kinetic term 
\begin{equation}\label{kinetic}
S_K= \frac{1}{4}\sum_{ij} \sigma_{ij} (X_i- X_j)^2 \,,
\end{equation}
and rewrite $\mathcal{P}_{\bar\Gamma}$ above as an interaction potential 
\begin{equation}\label{potential}
S_I[\xi,\theta]=-\sum \log(i U(\theta_i))-i \xi_i \theta_i + f\left( \sigma_{ij}^{\frac{d}{2}-1}\xi_i \xi_j \right) \,,
\end{equation}
where the function $f$ is 
\begin{equation}
f\left( \sigma_{ij}^{\frac{d-2}{2}}\xi_i \xi_j \right)= \log( \sum_ m \tfrac{\Gamma\left(\frac{d-2}{2}\right)^m(-1)^m}{2^{\frac{m d}{2}} \Gamma \left(  m \left( \frac{d -2}{2} \right) \right) } \sigma_{ij}^{m \left(  \frac{d- 2 }{2} \right)} (\xi_i \xi_j)^m ).
\end{equation}
and the extra factor of $\frac{1}{\sigma_{ij}}$ has been absorbed into the measure for $\sigma$.
To obtain the expression (\ref{potential}) we followed the trick
$$ \left(\cdots\right)\bigg|_{\xi=0} = \int_{-\infty}^{\infty} d\xi \int_{-\infty}^{\infty} d\theta \,e^{-i \xi\theta} \left(\cdots\right)\, ,$$
and performed integration by parts inside the path integral. 

\section{Triangulation and continuum}
We view the sum of the kinetic and potential terms in (\ref{kinetic}) and (\ref{potential}) as discretized worldsheet action of the dual string theory and take the continuum limit to obtain the continuous worldsheet theory. This should be done with care employing the technology of Strebel differentials and Discrete Exterior calculus which we shortly review below, and in more detail in the Supplementary material. 
\subsection{Strebel differentials}
Theory of Strebel differentials \cite{mulase1998ribbon} provides an isomorphism between the space of metric ribboned graphs and the moduli space on punctured Riemann surfaces with a positive real number assigned to every puncture. To every propagator with length $\sigma$ corresponds a vertical strip in the complex plane of width $\sigma$ \cite{Gopakumar:2005fx,Gopakumar:2022djw}. Then, these strips are glued into a manifold as depicted in figure \ref{fig:worldsheet} and detailed in the Supplementary material. The vertices of $\Gamma$ become punctures in the manifold and a positive real number $a_i$ associated to the $i$-th puncture is given by the sum of all the lengths of edges incident to the corresponding vertex $a_i = \sum_{j}\sigma_{ij}$. 
\begin{figure}
	\centering
	\includegraphics[width=\linewidth]{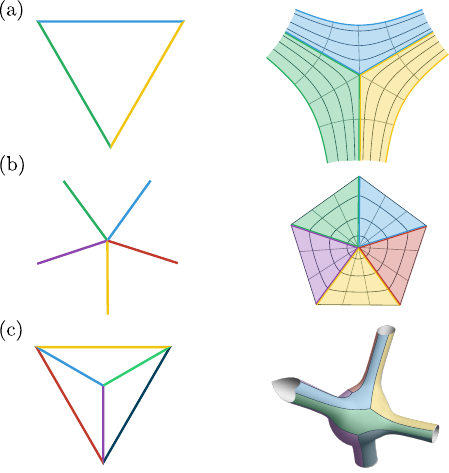}
	\caption{Gluing of the strips at (a) a face and (b) a vertex. In (c) we show an example of the Riemann surface obtained from a simple graph of genus $0$ with $4$ vertices corresponding to a sphere with $4$ infinite cylinders attached.}
	\label{fig:worldsheet}
\end{figure}
This construction naturally provides a metric for the Riemann surface given by the flat metric on each strip, corresponding to a particular gauge fixing of Weyl and diffeomorphism symmetries. This metric exhibits a double pole around the punctures (vertices of $\Gamma$)
\begin{equation}\label{eq::metpoles}
g_{Str}=  \frac{a_i^2}{4 \pi^2} \frac{d u d \bar{u}}{\left|u\right|^2}\, .
\end{equation}

\subsection{Discrete Exterior Calculus}
\label{sec:discrete_exterior_calculus}
Theory of Strebel differentials provide the worldsheet as a Riemann surface but it does not determine the worldsheet action from (\ref{kinetic}) and (\ref{potential}). For this, one first needs to extend the map between double poles of the Riemann surface (\ref{eq::metpoles}) and location of vertices in space time, $X_i^\mu$, to entire functions $X^\mu(u,\bar u)$ on the Riemann surface. We can address this issue through the machinery of Discrete Exterior Calculus used to solve geometric problems on manifolds through numerical methods where the specified data is necessarily finite \cite{desbrun2005discrete}. 

Consider a Riemann surface $\Sigma$ with a triangulation, see below, in terms of points $v_i$, intervals $e_{ij}$ and triangles $f_{ijk}$ such that the edges of $f_{ijk}$ are $e_{ij}$, $e_{jk}$ and $e_{ki}$ and the endpoints of $e_{ij}$ are $v_i$ and $v_j$. One discretizes  data on the manifold by integrating over the relevant simplexes of our triangulation. That is, one integrates a function over the points $v_i$, a 1-form over the intervals $e_{ij}$ and a 2-form over the triangles $f_{ijk}$. This allows us to use Stokes' theorem without issue. For instance, take the $1$-form $d f$ specified over the edges $e_{ij}$, then one has
\begin{equation}
	(d f)_{e_{ij}}=\int_{e_{ij}} d f = \int_{\partial e_{ij}} f= (f)_j- (f)_i\, .
\end{equation}
This defines for us a discrete exterior derivative which is unambiguous in the discretization lengths by virtue of Stokes' theorem.

\begin{figure}
	\centering
	\includegraphics[width=\linewidth]{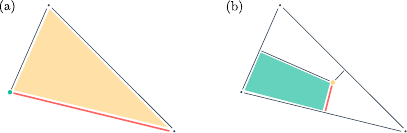}
	\caption{Depiction of part of the triangulation (a) with its dual shown in (b). Cells painted in the same color are dual to each other.}
	\label{fig:dual_triangulation}
\end{figure}

Another crucial operation is the Hodge dual. We first define the dual $\star \sigma^k$ of a $k$-simplex $\sigma^k$ by the convex span of the circumcenters of all the simplexes that contain $\sigma^k$ and $\sigma^k$ itself, see Figure \ref{fig:dual_triangulation}. 
The hodge dual of a $k$-form $\alpha$ is then defined by
\begin{equation}
\label{hodge_dual}
(\star \alpha)_{\star \sigma^k}=\frac{\left|\star \sigma^ k\right|}{\left|\sigma^k\right|}(\alpha)_{\sigma^k}
\end{equation}
where $\left|\cdot\right|$ stands for the volume in the corresponding dimension \footnote{Note that $\star \alpha$ is not discretized in the same triangulation as $\alpha$, but instead in a dual version of it.}. The last operation we need is the wedge product \cite{desbrun2005discrete}
\begin{equation}
(\alpha\wedge \beta)_{\sigma^k\cup \star \sigma^k} = \frac{\left|\sigma^k\cup \star \sigma^k\right|}{\left|\sigma^k\right| \left|\star \sigma^k\right|} (\alpha)_{\sigma^k} (\beta)_ {\star \sigma^k}
\end{equation}
where $\sigma^k\cup \star \sigma^k$ is the smallest convex set containing both $\sigma^k$ and $\star \sigma^k$. If we decrease the size of our triangulation we can expect all these operations to converge \cite{mohamed2018numerical,schulz2020convergence} since surfaces are locally flat and thus well approximated by subsets of $\mathbb{R}^2$.
\subsection{Triangulation}
\label{subsec:trinagulation}
\begin{figure}
	\centering
	\includegraphics[width=\linewidth]{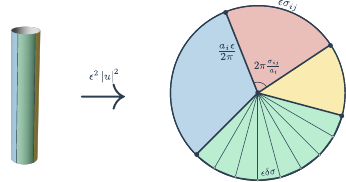}
	\caption{After we perform a Weyl transformation the infinite cylinders become flat disks split into sections that are identified along its curved edge. We triangulate the manifold by splitting this sections into small triangles.}
	\label{fig:Weyltransformation}
\end{figure}

Finally, we need to provide a triangulation of the Riemann surface---which we obtained using Strebel differentials above---on which discrete exterior calculus can operate. This is tricky, as the punctures are located at the endpoints of semi-infinite cylinders, see Fig.~\ref{fig:worldsheet}. We can, however, remove these cylinders by a Weyl transformation with a factor of $\epsilon^2 \left|u\right|^2$ in the region $0<\left|u\right|<1$ on the Strebel metric (\ref{eq::metpoles}). 
This leaves a flat punctured disk which can be completed by restoring the marked point. Disks corresponding to different poles are identified at the horizontal curves at $\left|u\right|=1$ defining a continuous Weyl transformation on the entire manifold. 
The resulting manifold is a collection of sections of flat disks with opening angles $\frac{2 \pi \sigma_{ij}}{a_i}$ and radii $\frac{a_i \epsilon}{2 \pi}$ identified along their curved edges, see Fig.~\ref{fig:Weyltransformation}. The factor of $\epsilon$ enables us to adjust the volume $V$ of the manifold, i.e. to keep it fixed even when we introduce more poles or change the $a_i$. One finds 
\begin{equation}
 V= \sum_i \frac{\epsilon^2 a_i^2}{4 \pi}\qquad \Rightarrow\qquad \epsilon= \sqrt{\frac{4 \pi V}{\sum a_i^2}}\sim \frac{1}{\sqrt{n}}\, ,
 \end{equation}
What we achieved so far is not quite a triangulation because sections of discs are not simplexes, but we can split each section further into small triangles that approximate the discs arbitrarily well. Note that at strong coupling, where diagrams with $n\gg 1$ dominate the partition function, the mesh describing our manifold becomes finer and finer. 


We will need the following geometric data, which we can easily deduce from Fig.~\ref{fig:Weyltransformation}. Denoting the poles by indices $i$ and $j$ and the auxiliary vertices introduced to make the discs into triangles by an index $a$, we obtain $\left|e_{ia}\right|=a_i \epsilon/2 \pi$ for length of the auxiliary edges and $\left|\star e_{ia}\right|=\epsilon \delta \sigma/2$ for their duals. Their convex span has area $\left|e_{ia}\cup \star e_{ia}\right|= a_i \delta \sigma \epsilon^2/8 \pi$. The area of the dual of a pole is $\left|\star v_i\right|= a_i^2 \epsilon^2/16 \pi$ and for the auxiliary vertices lying between disk $i$ and disk $j$ we find $\left|\star v_{a}\right|= 3(a_i+a_j) \delta \sigma \epsilon^2/16 \pi$.
\subsection{Continuum}

We are ready to take the continuum limit of the discrete action (\ref{kinetic}) and (\ref{potential}). Discrete exterior derivative of $X^\mu$ is given by
\begin{equation}\label{eq:dX}
(\dd X^\mu)_{e_{i a}} = \frac{a_i}{a_i +a_j} \left( X_j-X_i \right) 
\end{equation}
where the vertex $v_{a}$ lies between the disks $i$ and $j$ and we chose to linearly interpolate the known values of $X^\mu$ to the auxiliary vertices on the edges of the disk. Following the definitions above one finds
\begin{equation}
(a \dd X^\mu \wedge \star \dd X_\mu)_{e_{ia}\cup\star e_{ia}+e_{ja}\cup\star e_{ja}}=\frac{\pi \delta \sigma}{4}(X_j-X_i)^2
\end{equation}
Summing over the entire manifold yields 
\begin{equation}
\label{eq:kineticterm}
S_K= \frac{1}{\pi} \int d^2 z \sqrt{\det g} \, a\,\nabla_ \alpha X^\mu \nabla_ \beta X^ \nu g^{\alpha \beta} \eta_{\mu \nu}\, ,
\end{equation}
up to corrections of order $\epsilon\sim 1/\sqrt{n}$.

For the continuum limit of the potential term (\ref{potential}) we need to find a continuum version of the $\sigma$s. There are generically $6g- 6 + 3n$ independent $\sigma$s due to the correspondence with the space $\mathcal{M}_{g,n}\times \mathbb{R}_+^n$. A total of $n$ of them are expressible in terms of $a$ and $6g- 6 + 2 J$ where $J$ is the number of external sources are still present in the stringy calculations as moduli of the worldsheet. This leaves $2(n-J)$ to be integrated out which can be rearranged into two more real functions $t_1 $ and $t_2$. We define these to be given by the interpolation of $\frac{\sigma_{ij}}{a_i}$ for the independent $\sigma_{ij}$ which are not one of the $6g-6+2J$ moduli of the worldsheet. At the external sources where there cannot be any independent $\sigma$ we just set $t_i$ to $0$. There are still some $\sigma$ at every vertex that are not captured by these functions. This is either because they correspond to moduli of the worldsheet or because they can be expressed in terms of $a$. The first kind we will ignore for the moment since they are clearly subleading in $\frac{1}{n}$. For the second kind we note that a typical skeleton vertex has order $3$ unless some of the $\sigma$ are fine tuned to $0$, thus we can construct a third function given by $t_3= 1-t_1-t_2$ to capture them. If vertices do not have valence $3$ we can include auxiliary vertices and propagators with vanishing length to generate higher valence vertices. Doing this will restrict the integral over the moduli to a submanifold of moduli space. The other quantity in (\ref{potential}) for which we need a continuum approximation is $\xi_i \xi_j$. To make sense of the delta functions in $f_p$ let us assume that we smear them in a rotationally symmetric fashion on the region dual to $v_i$. Then it is straightforward to show that $\left(\frac{\dd \xi \wedge \star \dd \xi}{\star n f_p}\right)_{\tilde{v}_{ia}}= \frac{\pi}{2} (\xi_i -\xi_a)^2$ by just dividing the values of the two top forms integrated on $\star \tilde{v}_{ia}\equiv e_{ia}\cup \star e_{ia}$ and similarly $\left(\frac{\dd a\wedge \star \dd \xi}{ \star n f_p}\right)_{\tilde{v}_{ia}}= \frac{\pi}{2} \frac{a_i^2 (a_i-a_j)(\xi_i -\xi_j)}{(a_i+a_j)^2}$. Interpolating these to $v_a$ we directly find
\begin{equation}
\xi_i \xi_j = \left(\xi^2 + \frac{4 \xi}{a \pi}  \frac{\nabla \xi\cdot \nabla a}{n f_p}- \frac{2}{ \pi} \frac{(\nabla \xi)^2}{n f_p}\right)_{v_a}+ \mathcal{O}\left( \frac{1}{n^{\frac{3}{2}}} \right) \, .
\end{equation}
Using the definition of $f_p$ to turn (\ref{potential}) into an integral we can express its continuum limit as 
\begin{align}
\label{eq:interactionterm}
&S_I=- n\int \left(\log(i U(a^{\frac{d-2}{4}}\theta))- i \xi \theta + \sum_i f\left(t_i^{\frac{d-2}{2}}\xi^2 \right)\right )\star f_p  \nonumber\\
&\phantom{=}  -\int C_1\left( \frac{4}{\pi} \nabla \log \xi a^{-\frac{d-2}{4}}\cdot \nabla \log a- \frac{2}{\pi}(\nabla \log\xi a^{-\frac{d-2}{4}})^2\right)  \star 1 
\end{align}
where we have rescaled $\xi\to \xi a^{-\frac{d-2}{4}}$ and $\theta \to \theta a^{\frac{d-2}{4}}$ and defined $C_1=\sum_i\xi^2 t_i^{\frac{d-2}{2}} f'(\xi^2 t_i^{\frac{d-2}{2}}) $. For more details of this computation we refer to \cite{followup}. 

Putting everything together we find the following expression for the generating function 
\begin{equation}
\label{eq:pathintegral}
\mathcal{Z}[J]= \sum_n \frac{1}{n!}\int \mathcal{D}X^\mu \mathcal{D}\xi \mathcal{D}\theta \mathcal{D} a \mathcal{D}[g] \left|\frac{\delta \sigma}{\delta (a,g) } \right| e^{-S_K-S_I}
\end{equation}
where $S_K$ and $S_I$ are given by \eqref{eq:kineticterm} and \eqref{eq:interactionterm}. The path integrals are understood to be discretized on the triangulation described above and $\mathcal{D}[g]$ refers to an integral over the moduli space of the Riemann surfaces with marked points. 

Leaving the Jacobian  $\left|\delta \sigma/\delta (a,g) \right|$ aside for a moment, we now consider the large $n$ limit of (\ref{eq:pathintegral}). Clearly the path integrals over $\xi$, $\theta$, $t_1$ and $t_2$ will be dominated by the classical saddle of (\ref{eq:interactionterm}). Note that for the leading term on the classical saddle the region of integration of the moduli becomes irrelevant and it only contributes to the 1-loop determinant.
These can be further be solved in powers of $a$ around $a\to\infty$ which is also a natural limit to take, as it corresponds to the UV limit of the QFT, which we expect to be a CFT---see Discussion below. Restricting to the case of real fields with a single trace potential given by $U= \Tr\left(\sum \frac{\lambda_k}{k!} \Phi^k + J^T \Phi\right)$, we find that $\xi$, $\theta$, $t_1$ and $t_2$ are constant to leading order in $a$
\footnote{This holds only for $d\ge3$ otherwise the terms that dominate the contribution from the potential are no longer the highest order interactions.}. More general situations can be considered just as easily with similar results. Using this solution in \eqref{eq:interactionterm} we obtain
\begin{equation}\label{SIlargen}
S_I= C_0 + \frac{k(d+6)(d-2)}{16 \pi}\int d^2 z \sqrt{\det g} \frac{\nabla_ \alpha a \nabla^\alpha a}{a^2} 
\end{equation}
for some constant $C_0$ and suppressing a term proportional to $\sum_i \log a_i$ which we will take into account as part of the measure. 

Coming back to the measure in terms of $\sigma$ it is $\prod \frac{\dd \sigma_{ij}}{\sigma_{ij}}$. Translated into our choice of moduli and $a_i$, this gives a measure for the stringy moduli given by $\frac{\dd t_i}{t_i}$ and for $a_i$ we have $\frac{\dd a_i}{a_i}$. There is another factor for the moduli of the internal vertices which is irrelevant due to the saddle point approximation performed on $t_1$ and $t_2$. This will essentially determine $C_0$. The measure for the stringy moduli turns out to coincide with the stringy measure expected from the Fadeev-Popov determinant computed by the ghost sector of the string as can be directly checked. As for the metric for $a$, in a string theory we would expect a measure induced by the distance
\begin{equation}
    |\delta a|^2= \int\dd ^2 z \sqrt{\det g} \frac{\delta a^2}{a^2}\propto \sum_i \delta a_i^2\,.
\end{equation}
the difference between this measure and the one actually arising from the field theory is expressible as a Weyl transformation of the metric. Thus we will get some Liouville term for $\log a$. One can similarly show that the measure for $X^\mu$ suffers from the same exact problem giving more contributions to the Liouville term for $\log a$. Overall we find an extra contribution to the action given by
\begin{equation}\label{SLs}
    S_L=\int\dd^2 z\sqrt{\det g} K_1 \frac{(\nabla a)^2}{a^2}+ K_2 R \log a
\end{equation}
with $K_1=-\frac{16(1+9d)+8(d-2)k+(d-2)^2k^2}{1536\pi}$ and $K_2=-\frac{2 k - (k-12) d -4}{192 \pi}$. And the full stringy action is given by $S= S_K+ S_I+S_L$. This constitutes our final result for the worldsheet action in the large $n$ and $a\to\infty$ limits. We find an AdS$_{d+1}$ sigma model $ds^2 = du^2 + e^{-2u}\eta_{\mu\nu} dX^{\mu} dX^{\nu}$ with $a(z) = \exp(-2u(z))$ near the boundary. Deviations from this limit relate geometry toward the interior to the RG flow of the QFT. We leave study of such deviations to future work. 

\section{Discussion}
\label{sec::discussion}

We employed scalar matrix theories to illustrate a mechanism for emergence of string world-sheet from multi-loop Feynman graphs and demonstrated two basic features: (i) the holographic dimension arises from a continuum limit of Schwinger parameters, (ii) open-closed duality is realized by representing Strebel differentials in two equivalent ways, see fig. 2. There are possible limitations in our derivation. It might happen, for example, that the large $n$ limit of the perturbative sector we consider here corresponds to a sub-dominant contribution in the generating function. This, in principle, can be checked by standard resurgence techniques, see e.g. \cite{Mari_o_2014}, by performing Borel resummation followed by inverse Laplace transform. Then, one cannot make a sharp claim about the gauge-string duality even though our derivation of the world-sheet action, Eqs. (\ref{SIlargen}) and (\ref{SLs}), still holds. 
%
%
%
%
Second, we ignored issues of renormalization, which can systematically be carried out in the Schwinger formalism, see \cite{Itzykson:1980rh}, and assumed vanishing mass\footnote{See \cite{Gallegos:2022hbt} for inclusion of the mass term.}. One can demonstrate \cite{Gallegos:2022hbt,followup} that renormalization does not alter our main conclusions as long as the QFT possesses a critical limit $n\to\infty$, $\lambda \to \lambda_c$\footnote{Limiting cases $\lambda_c = 0$ and $\infty$ can be incorporated in the analysis.}. For example \cite{Zamolodchikov:1980mb} considered so-called ``fishnet'' theories and showed that this limit exists where the theory becomes a CFT for $(d,k)=(3,6)$, $(4,4)$ and $(6,3)$, precisely the combinations argued to possess a continuum limit in \cite{Gallegos:2022hbt}. Here the large vertex limit of the perturbative sector is also the dominant contribution hence avoiding the possible limitation discussed above.  In \cite{Basso:2018agi} the $(4,4)$ Fishnet theory has been shown to lead to string propagation in AdS$_5$ using integrability techniques. Our derivation provides an alternative without resorting to integrability. It is interesting to employ our approach to explore the Fishnet and Fishchain theories in more detail \cite{Gromov:2019bsj}, as well as connection to other studies \cite{Bhat:2021dez}. 

What happens beyond the CFT limit is of utmost importance. This can be obtained by going beyond the large $a$ limit in the generic expressions (\ref{eq:kineticterm}), (\ref{eq:interactionterm}). Furthermore, it is straightforward, albeit cumbersome, to include fermions and gauge fields in our derivation see for example \cite{Lam:1969xk}. Further generalizing our method to include renormalization properly, one could, in principle be able to derive the worldsheet action dual to QCD! 

Finally, we note that our computation is valid for arbitrary genus. In particular, (\ref{eq:pathintegral}) holds even though the Strebel metric varies yet remaining unique. It is tempting to ask whether we can extend our analysis to non-perturbative quantum gravity. In principle, our starting point, eqs.~(\ref{kinetic}) and (\ref{potential}) remains valid for arbitrarily small $N$. See \cite{Gallegos:2022hbt} for a discussion on this issue. 

There are various other exciting questions to explore in the future, for instance string propagation on blackholes through  extending our analysis to thermal field theory and dynamical gravity through, perhaps, generalizing to ensemble of QFTs.      

\section{Acknowledgments}

We are grateful to Rajesh Gopakumar, Edward Mazenc and Claire Moran for discussions. This work was supported, in part, by the Netherlands Organisation for Scientific Research (NWO) under the VICI grant VI.C.202.104.
\appendix
\renewcommand\thefigure{\thesection.\arabic{figure}}    
\section{Strebel Differentials}
\setcounter{figure}{0} 

\begin{figure}
	\centering
	\includegraphics[width=0.75\linewidth]{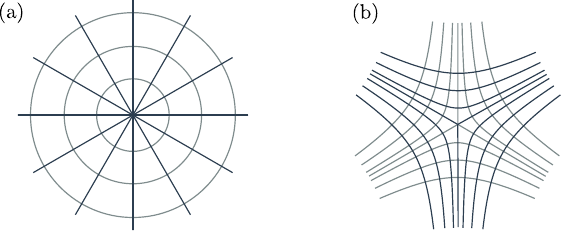}
	\caption{Structure of a Strebel differential near a double pole (a) and a single zero (b). The darker lines correspond to vertical curves and the lighter ones to horizontal curves.}
	\label{fig:zeroesandpoles}
\end{figure}

\label{app:Strebel}
Let $\Sigma$ be a Riemann surface of genus $g$ with $n$ marked points $\left\{ p_1,\dots, p_n \right\}$ equipped with a complex structure. The basic objects we are interested in are quadratic meromorphic forms, that is, objects $\chi\in\Omega^{(1,0)}_1(\Sigma)\otimes \Omega^{(1,0)}_1(\Sigma)$ which in holomorphic coordinates locally take the form
\begin{equation}
	\chi= \phi(z) \dd z^2
\end{equation}
with $\phi(z)$ being a holomorphic function that transforms as $\phi(w)= \left(\frac{\dd w}{\dd z }\right)^2 \phi(z)$ under holomorphic changes of coordinates. Using these differentials one can define a special set of curves $\gamma$ such that $\chi(\dot{\gamma},\dot{\gamma})>0$. We call such curves horizontal since for $\Sigma=\mathcal{C}$ with the usual metric and choosing $\chi=\dd z^2$ they correspond to lines of constant imaginary part. We similarly denote by vertical curves those for which $\chi(\dot{\gamma},\dot{\gamma})<0$. At regular points of $\phi$ the vertical and horizontal curves form a grid which specify a set of coordinates for which $\chi= \dd z^2$. Near zeroes or poles of $\phi$ this is no longer true and we depict the local structure of vertical and horizontal lines in Figure \ref{fig:zeroesandpoles}. Note that from a quadratic differential one can define a metric for the manifold given by
\begin{equation}
\label{eq:strebelmetric}
  g_{Str}= |\phi(z)|\dd z \dd \bar{z}
  \end{equation}
In the special coordinates defined by vertical and horizontal curves this metric reduces to patches of the Euclidean metric in $\mathcal{C}$ along with curvature localized at the zeroes and poles of $\phi$.

The main result of the theory of Strebel differentials is given by the following theorem
\begin{theorem}
\label{th:Strebel}
Given a Riemann surface $\Sigma$ of genus $g$ with $n$ marked points $\left\{ p_1,\dots, p_n \right\}$ and a set of $n$ positive real numbers $\left\{ a_1,\dots, a_n \right\}$ one can find a unique quadratic meromorphic form $\chi$ such that
\begin{itemize}
\item $\chi$ locally takes the form $\phi(z) \dd z^2$ with $\phi(z)$ holomorphic on $M/\left\{ p_1,\dots,p_n \right\}$
\item $\phi(z)$ has a double pole at every $p_i$
\item The union of all non-compact horizontal curves is a set of measure $0$
\item Every compact horizontal curve $\gamma$ is a simple loop around some $p_i$ and it satisfies
\begin{equation}
a_i=\int_\gamma \sqrt{\phi(z)} \dd z
\end{equation}
\end{itemize}
\end{theorem}
Such a quadratic form is called a Strebel differential and the set of non-compact horizontal curves is called its critical graph. These corresponds to horizontal lines connecting two zeroes such that the critical graph can be regarded as the embedding of some abstract graph in our manifold.

In light of theorem \ref{th:Strebel} one can deduce that Strebel differentials parametrize the space $\mathcal{M}_{g,n}\times \mathbb{R}^n_+$ where $\mathcal{M}_{g,n}$ is the moduli space of Riemann surfaces of genus $g$ with $n$ punctures and one can take $g_{Str}$ as a representative of the classes of metrics up to Weyl and diffeomorphism transformations for some fixed collection of residues at every marked point.

From a metric ribboned graph $\Gamma$ one can construct a Riemann surface with its corresponding Strebel differential such that its critical graph is given by $\Gamma^*$, that is, the dual graph of $\Gamma$ obtained by replacing vertices by faces and vice-versa. The construction goes as follows. 
Let $\Gamma$ be an arbitrary metric ribboned graph with all faces bounded by at least three distinct edges. For every edge $e_i$ of $\Gamma$ with length $\sigma_i$ let us take an infinite strip in $\mathbb{C}$ oriented along the imaginary axis and of width $\sigma_i$. On these strips $\dd z_i^2$ defines a holomorphic quadratic form. Take any vertex $v_i$. Since $\Gamma$ is ribboned we have an ordering of the edges incident to this vertex, then we can glue the corresponding strips into a cylinder according to this ordering. This can be done by mapping all these strips into a single disk according to the map
\begin{equation}
  u_i(z_k)= \exp \left( 2 \pi i \frac{z_k +\sum_{j< k} \sigma_j }{\sum_j \sigma_j} \right)
\end{equation}
Pulling back the quadratic differentials on the strips to the $u$-disk we obtain
\begin{equation}
\dd z_k^2 = - \frac{(\sum \sigma_j)^2}{4 \pi^2} \frac{\dd u_i^2}{u^2_i}
\end{equation}
so that the form develops a second order pole at the points $u_i=0$. Here $k$ runs over all the vertices incident to $v_i$ according to the ordering induced by the ribbon structure of $G$. We also do something similar for the faces. Take any face $f_i$ with $m\ge 3$ edges bounding it. We now glue the strips of the edges forming the loop into a disk by mapping them with
\begin{equation}
 w_i(z_k)= e^{\frac{2 \pi i k}{m}} z_k^{\frac{2}{m}}
 \end{equation}
Now the quadratic differentials are mapped to
\begin{equation}
\dd z_k^2= \frac{m^2}{4} w_i^{m-2} \dd w_i^2
\end{equation}
so that they develop a zero of order $m-2$ at the points $w_i=0$. The collection of the strips with coordinates $z_i$ for the edges, the disks with coordinates $u_i$ for the vertices and the disks with coordinates $w_i$ for the faces define an atlas with holomorphic transition functions given by $u_i(z_k)$ and $w_i(z_k)$. Thus they completely characterize a Riemann surface with the same genus as the graph. Moreover the form $\chi$ with local expression on the strips given by $\dd z_i^2$ is the unique Strebel differential with marked points given by $u_i=0$ and residues $\sum_{e_j\in v_i} \sigma_j\equiv a_i$. In Figure 1 of the main text we give a drawing of how the strips are glued at vertices and faces and an example of the resulting Riemann surface from a simple graph.

There is no obstacle to following the same construction for diagrams that have two-edge loops, however the strips corresponding to the two edges get glued together into a single strip without any zero and the information that they were originally separate is lost.

This construction is actually an isomorphism between the set of skeleton ribboned graphs with lengths assigned to each edge with $\mathcal{M}_{g,n}\times \mathbb{R}_+^n$ with the positive numbers $a_i$ being the coordinates on $\mathbb{R}_+^n$. Note additionally that we can construct a triangulation of our Riemann surface with $0$-cells given by the zeroes and the poles, $1$-cells given by the horizontal curves connecting zeroes to zeroes and the vertical curves connecting zeroes to poles and $2$-cells given by half-strips split at $\Im(z_i)=0$.

\bibliographystyle{apsrev4-1}
\bibliography{refs}

\end{document}